\documentclass[12pt]{iopart}

\usepackage{graphicx}
\usepackage[english]{babel}

\begin{document}

\title{Qualitative Approach to Semi-Classical Loop Quantum Cosmology}

\author{G.V. Vereshchagin}
\address {I.C.R.A. - International Center for Relativistic Astrophysics, \\
University of Rome "La Sapienza", Physics Department,
P.le A. Moro 5, 00185 Rome, Italy.}

\begin{abstract}
Recently the mechanism was found which allows avoidance of the cosmological singularity within the semi-classical formulation of Loop Quantum Gravity. Numerical studies show that the presence of self-interaction potential of the scalar field allows generation of initial conditions for successful slow-roll inflation. In this paper qualitative analysis of dynamical system, corresponding to cosmological equations of Loop Quantum Gravity is performed. The conclusion on singularity avoidance in positively curved cosmological models is confirmed. Two cases are considered, the massless (with flat potential) and massive scalar field. Explanation of initial conditions generation for inflation in models with massive scalar field is given. The bounce is discussed in models with zero spatial curvature and negative potentials.
\end{abstract}



\maketitle

\section{Introduction}

Over a long period of time cosmologists believed that quantum cosmology would find a way to treat initial as well as final singularity. Loop Quantum Gravity (LQG) is at present the main background independent and nonperturbative candidate for quantum theory of gravity \cite{Rov98,Thi03}. Loop Quantum Cosmology (LQC) is the application of LQG to an homogeneous minisuperspace environment. Calculations performed within LQG established the possibility to avoid singularity \cite{Boj01}. Semi-classical equations within LQC were studied recently and some intriguing conclusions were obtained. First of all, the Big Crunch does not appear in models with realistic effective potentials \cite{Sin04}. Moreover, if initial conditions are chosen at the bounce so that the scalar field rests at the minimum of the effective potential, during the subsequent evolution it climbs the potential and hence generates initial conditions for successful slow-roll inflation \cite{Boj03,Lid04}.

These results were obtained by numerical integration of LQC equations and although authors claim that their results are quite generic, application of alternative methods to the problem is of interest. Qualitative analysis is quite a powerful method and it was successfully applied within the framework of General Relativity (GR) (see e.g. \cite{Bel85,Top97}) and gauge theories of gravity \cite{Ver03}. In this paper qualitative analysis of semi-classical LQC equations, which are presented in the next section, is performed. The origin of bounce is discussed in section 3. Simple model of massless scalar field with absolutely flat potential is studied in section 4. Qualitative behavior of cosmological model with massive scalar field is considered in section 5. Conclusions follow in the last section.

\section{Semi-classical LQC equations}

Near the classical singularity the picture of continuous spacetime breaks down and evolution within LQC is described by difference equations. However there is a region where semi-classical differential equations are valid but quantum effects are already significant. In this region of small distances the geometrical density $a^{-3}$ acquires quantum corrections. Good approximation to eigenvalues of geometrical density operator in LQC for large quantization parameter $j$ is given by \cite{Boj02}
\begin{eqnarray}
d_j(a)=D(q)a^{-3},
\end{eqnarray}
where $q=(a/a_*)^2$, and $a_*^2=\frac{j\ln2}{3\sqrt{3}\pi}l_P^2$ is the scale where quantum corrections become essential. It can be larger than the planckian scale $l_P$, since the quantization parameter $j$, which must take half integer values, is arbitrary \cite{Sin04,Lid04}. The quantum correction factor $D(q)$ is given by
\begin{eqnarray}
D(q)=\left(\frac{8}{77}\right)^6q^{3/2}\left\{7\left[(q+1)^{11/4}-|q-1|^{11/4}\right]-\right. \\ \left.\quad\quad\quad\quad\quad\quad\quad\quad\quad
-11q\left[(q+1)^{7/4}-\mathrm{sign}(q-1)|q-1|^{7/4}\right]\right\}^6.
\end{eqnarray}
Above the scale $a_i=0.357l_P$ the discrete nature of spacetime becomes unimportant and semi-classical equations \cite{Boj03,Sin04,Lid04} are valid
\begin{eqnarray}
\label{eq1}
H^2+\frac{k}{a^2} &=& \frac{8\pi}{3M_P^2}\left[\frac{\psi^2}{2D}+V\right],\\
\label{eq2}
\quad \quad \quad \dot\psi &=& -3H\psi+\frac{\dot D}{D}\psi-D\frac{\partial V}{\partial\varphi}, \\
\label{eq3}
\dot H+H^2 &=& -\frac{8\pi}{3M_P^2}\,\frac{\psi^2}{D}\,\left[1-\frac{\dot D}{4HD}\right]+\frac{8\pi V}{3M_P^2},
\end{eqnarray}
where $V(\varphi)$ is the effective potential, $\psi\equiv\dot\varphi$ is the scalar field derivative with respect to time, $H=\frac{\dot a}{a}$ is the Hubble parameter, $M_P$ is the planckian mass, the dot denotes differentiation with respect to time, and $k$ gives the sign of spatial curvature. According to (\ref{eq1}) if $V\geq0$, since the function $D$ is nonnegative, the bounce is possible only for positively curved spatial geometry, exactly as within GR, and we assume $k=+1$ until the end of section 5.

The dynamical system which corresponds to cosmological equations (\ref{eq1}-\ref{eq3}) for $k=+1$ can be represented as follows:

\begin{eqnarray}
\label{Hdot}
\dot H &=& \frac{4\pi}{M_P^2}\,\frac{\psi^2}{D}\,\left[\frac{1}{3}\left(\frac{a}{a_*}\right)^2\frac{D'_q}{D}-1\right]+\frac{1}{a^2}, \\
\label{psidot}
\dot\psi &=& -3H\psi\left[1-\frac{1}{6}\left(\frac{a}{a_*}\right)^2\frac{D'_q}{D}\right]-D\frac{\partial V}{\partial\varphi}, \\
\dot a &=& Ha, \\
\dot\varphi &=& \psi,
\end{eqnarray}
where\footnote{In numerical calculations it is convenient to make interpolated tables for the function $D(q)$ and its derivative $D'_q$. In classical limit (for large $q$) the former approaches unity while the latter vanishes.} $D'_q=dD/dq=\dot D(a/a_*)^{-2}/(2H)$. This is four-dimensional system for dynamical variables $\{H,\psi,a,\varphi\}$. Notice immediately, that the scalar field influence on dynamics of the system is due to the unique (last) gradient term in (\ref{psidot}). When the effective potential is completely flat, $V=V_0=$const, this term vanishes. Moreover, additional simplification is possible here, since the scalar field derivative $\psi$ can be expressed from (\ref{eq1}) and substituted into (\ref{eq3}) thus reducing the dimension of the dynamical system to two. The system can be completely characterized in terms of $\{H,a\}$ in that case.

From the other hand, for many types of effective potentials the gradient of the potential can be expressed as a function of the potential itself. Consider massive scalar field with $V=\frac{m^2\varphi^2}{2}$, where $m$ is a mass, and rewrite the gradient term with the help of (\ref{eq1})
\begin{eqnarray}
\label{dVdf}
\frac{\partial V}{\partial\varphi}=\pm\sqrt{2V}=\pm m\sqrt{\frac{3M_P^2}{4\pi}\left(H^2+\frac{k}{a^2}\right)-\frac{\psi^2}{D}}.
\end{eqnarray}
This expression can be used in order to reduce the dimension of the dynemical system until it does not change sign. This is true for the whole evolution of the dynamical system during bounces and inflation, described e.g. in \cite{Lid04}. In that case the system can be reduced to three-dimensional one with the following dynamical variables $\{H,\psi,a\}$.

\section{Origin of bounce in semi-classical LQC}

\begin{figure}[ht]
\begin{center}
\includegraphics[width=5.5in]{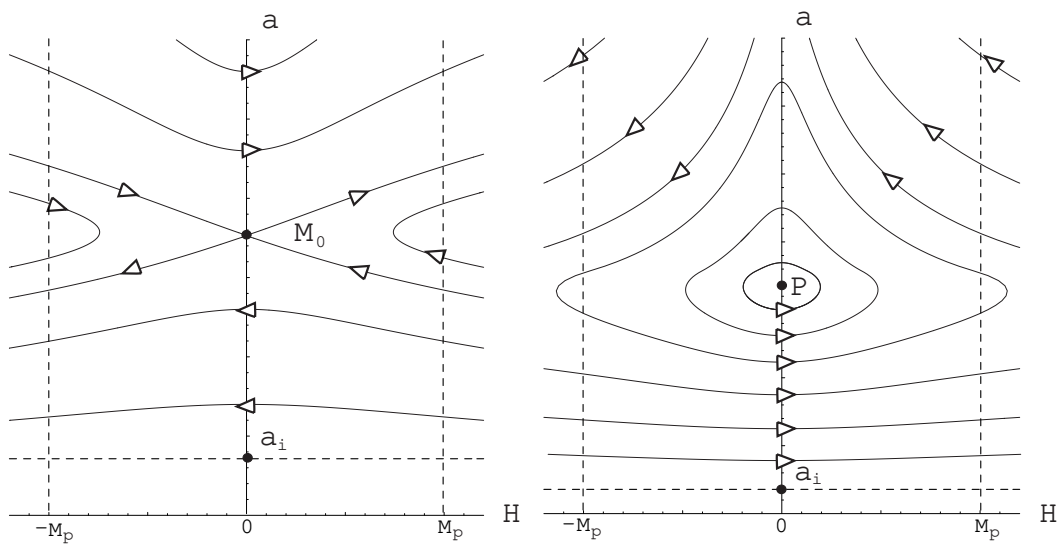}
\end{center}
\caption{Left figure: saddle singular point in the section of the phase space which guarantees the bounce. Here $j=30000$ just to visualize the saddle. For reasonably small quantization parameter $j<400$ this singular point lies below the scale of quantum geometry $a_i$. Right figure: focus singular point in the phase space for masslss scalar field. Phase trajectories are closed curves and the cosmological parameters experience oscillations. The quantization parameter is $j=300$. Dashed lines surround the region where semi-classical LQC is valid, i.e. where $|H|<M_p$ and $a>a_i$.}
\label{fig1}
\end{figure}

The possibility of bounce emerges within semi-classical LQC since the phase space differs essentially from the case of GR for small scale factors due to quantum corrections. The function
\begin{eqnarray}
\label{Qf}
Q(a)=\left(\frac{a}{a_*}\right)^2\frac{D'_q}{D}
\end{eqnarray}
can change sign in (\ref{Hdot}) so the Hubble parameter derivative may vanish. In the case of massless scalar field equality $Q=2$ determines the singular point of the dynamical system. The right hand side of equation (\ref{dH}) (see below) should be used in order to determine the singular point position with fixed value of the effective potential\footnote{We do not consider here all singular points of the system (\ref{eq1}-\ref{eq3}). For instance the usual focus at $\varphi=\psi=0$, which is responsible for oscillations of the scalar field after inflation, is ignored.}. It is the saddle singular point on the $a$ axis in the phase space that ensures the bounce for solutions of cosmological equations (see fig. \ref{fig1} left). One can see that the bounce is guaranteed for all phase trajectories crossing the $a$ axis above the saddle $M_0$, since the Hubble parameter changes sign from negative to positive and the transition from contraction to expansion takes place. The position of the saddle $M_0$ for massless scalar field depends on quantization parameter as
\begin{eqnarray}
\label{am0}
a_{M_0}\approx1.765\cdot10^{-2}\sqrt{j}.
\end{eqnarray}
With nonvanishing potential it depends on the value of the effective potential very weakly. Only for very large $j$ with large $V_0$ the position of the saddle is shifted a little towards smaller values.
Expression (\ref{am0}) shows that the saddle point comes into the region of validity of semi-classical LQC ($a>a_i$) for $j>400$. This means that trajectories laying below the point $M_0$ appear in the region of semi-classical LQC for $j>400$. These trajectories do not have bounce since they describe transition from expansion to contraction. However, these trajectories originate from and come back to the rigion below $a_i$ where purely quantum geometric effects should be responsible for the absence of singularity \cite{Boj01}.

Assume below that the saddle singular point $M_0$ lies below the scale $a_i$, or, equivalently, $V_0<M_P$ and $j<400$.

\section{Scalar field with flat potential}

Assume in this section that the effective potential of the scalar field is completely flat, i.e. $V=V_0$=const. Corresponding dynamical system is given by the following equations
\begin{eqnarray}
\label{dH}
\dot H &=& \left(H^2+\frac{k}{a^2}-\frac{8\pi V_0}{3M_P^2}\right)(Q-3)+\frac{k}{a^2}, \\
\dot a &=& Ha.
\end{eqnarray}

Consider first the case of purely massless scalar field $V_0=0$. The phase space is represented at fig. \ref{fig1} (right). The only singular point on the axis $a$ in addition to $M_0$ is the focus $P$. Its position dependence on the value of parameter $j$, determined empirically, is
\begin{eqnarray}
a_P\approx0.1944\sqrt j.
\end{eqnarray}
Therefore, $a_P>a_i$ for $j>3.5$. This means, oscillating behavior of cosmological models with massless scalar field found in \cite{Lid04} is very generic. From the other hand, $a_P>3.89$ when $j>400$, i.e. when the saddle of fig. \ref{fig1} enters the scale of semi-classical LQC. However, this does not change the qualitative picture of oscillations since the asymptotes of the saddle are almost horizontal for small $j$ and almost all phase trajectories experience a bounce.

\begin{figure}[ht]
\begin{center}
\includegraphics[width=6in]{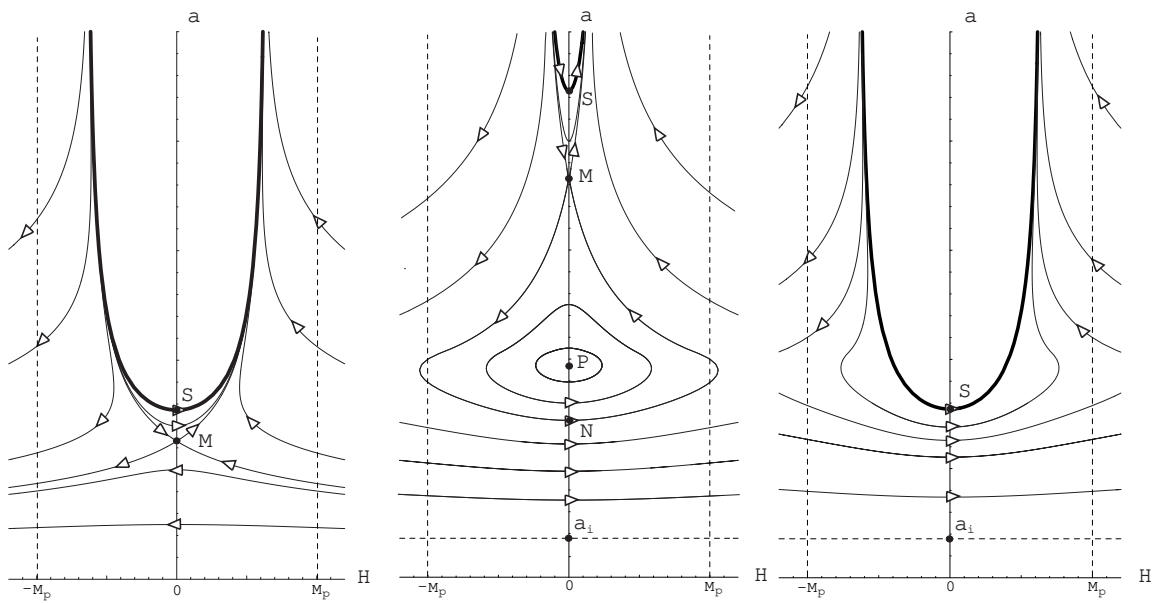}
\end{center}
\caption{Phase portraits of cosmological models with the scalar field with flat effective potential. Left figure corresponds to the case of GR, $V_0=0.05M_P$. Central figure represents the phase space with focus and saddle for $V_0=0.006M_P$. Right figure represents the phase space with $V_0=0.05M_P$. Thick curves surround regions where the derivative of the scalar field $\psi$ is complex and there are no solutions. For central and right figure $j=100$. Dashed lines again surround the region where semi-classical approach is valid.}
\label{fig2}
\end{figure}

Now assume the value of the potential is nonvanishing, $V_0>0$. Corresponding phase portraits are shown at fig \ref{fig2}.
Left figure represents the case of GR. The saddle $M$ here originates from the fact that $Q=2$ in (\ref{dH}) in classical limit and the bounce becomes possible for sufficiently large $V_0$. The thick curve surrounds the region where bounce is impossible since the scalar field derivative gets complex and solutions are absent. Phase trajectories corresponding to bouncing solutions can cross the $a$ axis only between points $S$ and $M$. At the point $S$ the derivative of the scalar field vanishes. Stability of solutions of this type were studied in \cite{Ver04}. It turns out that they are exponentially unstable.

The structure of the phase space depends on some critical value of the potential $V_c$, so if $V_0>V_c$ some singular points disappear. The value $V_c$ is empirically
\begin{eqnarray}
\label{Vc}
V_c=2.128j^{-1}.
\end{eqnarray}
Central figure describes the phase space for $V_0<V_c$. There are two singular points at the axis $a$, the focus $P$ and the saddle $M$. The presence of additional saddle $M_0$ below $a_i$ still guarantees the bounce for all solutions. There are two types of possible solutions: oscillating solutions, when the scale factor at the bounce $a_{min}$ lies between $P$ and $N$ (the point where the saddle separatrix bounces), or otherwise quasi-exponential bouncing solutions.

Right figure corresponds to $V_0>V_c$. Both $P$ and $M$ disappear and the bounce below $S$ now possible for all solutions. Notice the difference between left and right figures. Instability of bouncing solutions within GR is due to the fact that the saddle separatrices have de Sitter asymptotes like all solutions. Corresponding separatrices do not have de Sitter asymptotic within semi-classical LQC and the class of bouncing solutions is much wider.

\begin{figure}[ht]
\begin{center}
\includegraphics[width=6in]{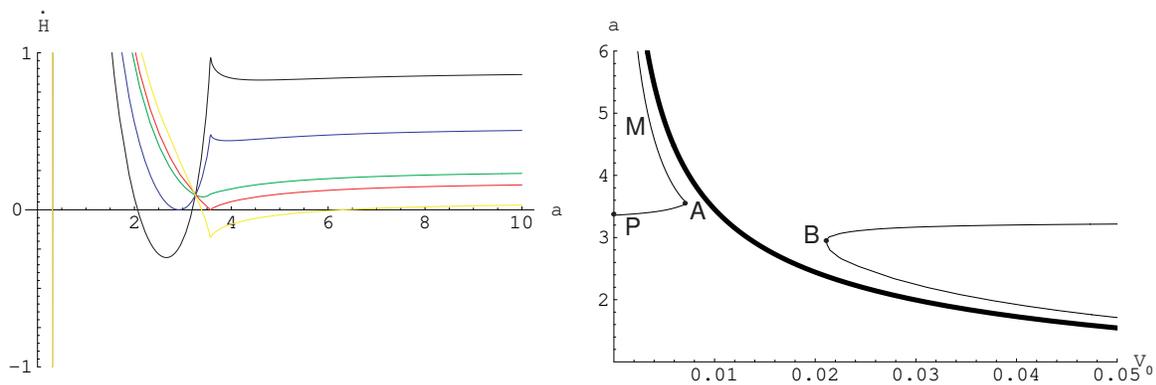}
\end{center}
\caption{Left figure: positions of singular points as functions of the scale factor for fixed values of $V_0$. There are three cases: the function $\dot H$ crosses the axis $a$ in two points (yellow and black curves), in one point (blue and red curves) and does not cross the $a$ axis (green curve). The effective potential grows from yellow to black curves. Right figure: dependence of singular points positions on the value of $V_0$. Above the thick curve $\psi$ is complex and there are no solutions. At the point $A$ the singular points disappear and at the point $B$ they appear again. The thick curve marks the scale factor values for which the scalar field derivative $\psi$ vanishes.}
\label{fig3}
\end{figure}

It is of interest to consider the dependence of positions of $M$ and $P$ on the values of $j$ and $V_0$. Notice, that although there is dependence on $j$, the qualitative results remain the same for a wide range of quantization parameter variation. Assume first $j=300$ and study only dependence of singular points positions on $V_0$. The result is shown at fig. \ref{fig3}.

Left figure here represents positions of singular points on axis $a$ dependence on the scale factor for fixed $V_0$. The potential has minimal value for yellow curve and grows monotonically for red, green, blue and black curves. There are three cases: two points $M$ and $P$ are present, only one point is left, no singular points are left. This situation is additionally illustrated on the right figure. For purely massless scalar field $V_0=0$, the point $M$ goes to infinity and we come to the phase portrait represented at fig. \ref{fig1} (right). For small effective potentials both singular points appear in the phase space. The position of $P$ depends on $V_0$ rather weakly while the position of $M$ is very sensitive to it. Above the thick curve the scalar field derivative becomes complex. Although the two singular points come to the phase space for large $V_0$ again, they lie in the region of complex $\psi$.

\section{Massive scalar field}

Now we are ready to describe the behavior of cosmological models with massive scalar field within semi-classical LQC. Although the phase space is four-dimensional, as mentioned at the end of section 2, we can use (\ref{dVdf}) to reduce its dimension to three until the oscillations of the scalar field after inflation are not considered.

Since the bounce is ensured by the presence of the saddle $M_0$ below the quantum geometry scale $a_i$ it is convenient to consider initial conditions at the bounce where $H=0$. From the results of the previous section it follows that the picture of bounce is completely determined by the value of the scale factor and effective potential at the bounce. When $V<V_c$, where $V_c$ is given by (\ref{Vc}) there are two singular points in the section of the phase space $\{H,a\}$ and oscillations develop if the scale factor lies between the points $P$ and $N$. However, during oscillations the value of the scalar field increases (if $\psi>0$) or decreases (when $\psi<0$) and the scalar field climbs the potential walls \cite{Lid04}. So in the section $\{H,a\}$ of the phase space the singular points $P$ and $M$ tend to each other until the value of the potential reaches the critical value $V_c$. When $V=V_c$ possibility of oscillations disappear and the phase trajectory inevitably escapes from the bounded region near the point $P$. The scale factor starts to grow quasi-exponentially, and the solution approaches classical region so inflationary regime of GR develops. When inflationary expansion begins the kinetic energy of the scalar field becomes suppressed and the subsequent expansion is determined by slow-roll of the scalar field. Notice that the less is the scalar field mass (the more flat is the potential), the more oscillations can develop before the inflationary stage begins. When $V>V_c$ there are no oscillations and the bounce is unique.

The same qualitative behavior should take place for all symmetric nonnegative effective potentials of the scalar field. It is interesting to note, that the possibility of bounce for zero curvature models ($k=0$) appears in semi-classical LQC with negative potentials. The phase portrait of zero curvature model with $V_0<0$ is similar to fig. \ref{fig1} (right).

Notice at last, that the validity of the phase analisys presented here is limited. For $V>M_P$ for instance the quantum nature of the scalar field becomes important. Also the Hubble parameter is constrained by the usual condition $H<M_P$. Finally, the requirement $a>a_i$ should be satisfied.

\section{Conclusions}

Qualitative study of dynamical system corresponding to semi-classical LQC equations is performed.
It is shown, that the bouncing character of cosmological solutions within semi-classical LQC is guaranteed by the presence of the saddle singular point on the scale factor axis. The bounce takes place for all solutions already in semi-classical region of LQC if $j<400$. Although some `singular' solutions appear in semi-classical region for $j>400$ they should be free of singularity due to quantum descrete structure of spacetime \cite{Boj01} which is not considered here. The possibility of oscillations before the bounce in models with massive scalar field is explained. It is shown that initially oscillating solutions inevitably followed by the inflationary stage. The bounce is possible also for zero curvature models within semi-classical LQC with negative potentials.

\section*{Acknowledgement} I am grateful to anonymous referee for important remarks. I would like also to thank Parampreet Singh for useful discussions.

\end{document}